\documentclass[12pt]{article}
\usepackage{amssymb,amsmath}
\usepackage{cite}
\numberwithin{equation}{section}

\newcommand{\bea}{\begin{eqnarray}}
\newcommand{\eea}{\end{eqnarray}}
\newcommand{\beano}{\begin{eqnarray*}}
\newcommand{\eeano}{\end{eqnarray*}}
\newcommand{\nonu}{\nonumber \\}

\newcommand{\hs}[1]{\hspace{#1 mm}}
\newcommand{\eps}{\varepsilon}

\newcommand{\Phidag}{\Phi^\dagger}

\newcommand{\adag}{a^\dagger}

\newcommand{\cd}{\mbox{$\cal{D}$}}

\newcommand{{\cg}}{\mbox{$\cal{G}$}}
\newcommand{\ch}{\mbox{$\cal{H}$}}

\newcommand{\cp}{\mbox{$\cal{P}$}}

\newcommand{\cs}{\mbox{$\cal{S}$}}
\newcommand{\calr}{\mbox{$\cal{R}$}}
\newcommand{\ct}{\mbox{$\cal{T}$}}

\newcommand{\cds}{\mbox{$\cal{D}_{\cal{S}}$}}
\newcommand{\cdst}{\mbox{$\cal{D}_{\widetilde{\cal{S}}}$}}
\newcommand{\prt}{\partial}

\newcommand{\wh}[1]{\widehat{#1}}
\newcommand{\wt}[1]{\widetilde{#1}}
\newcommand{\mb}[1]{\hs{4}\mbox{#1}\hs{4}}

\newtheorem{theo}{Theorem}[section]

\newtheorem{lem}[theo]{Lemma}

\newcommand{\prf}{\underline{Proof:}\ }
\newcommand{\finprf}{\null \hfill {\rule{5pt}{5pt}}\\ \null}
\newcommand{\ie}{{\it i.e.}\ }
\newcommand{\CC}{\mbox{${\mathbb C}$}}
\newcommand{\RR}{\mbox{${\mathbb R}$}}

\newcommand{\ZZ}{\mbox{${\mathbb Z}$}}

\newcommand{\II}{\mbox{${\mathbb I}$}}

\newcommand{\EE}{\mbox{${\mathbb E}$}}
\newcommand{\FF}{\mbox{${\mathbb F}$}}
\newcommand{\GG}{\mbox{${\mathbb G}$}}

\newcommand{\PR}[1]{Phys.\ Rev.\ {\bf #1}}

\begin{document}
\renewcommand{\thefootnote}{\fnsymbol{footnote}}
\newpage
\pagestyle{empty} \setcounter{page}{0}

\markright{\today\dotfill DRAFT\dotfill }


\newcommand{\LAP}{LAPTH}
\def\logo{{\bf {\huge LAPTH}}}

\centerline{\logo}

\vspace {.3cm}

\centerline{{\bf{\it\Large Laboratoire d'Annecy-le-Vieux de
Physique Th\'eorique}}}

\centerline{\rule{12cm}{.42mm}}

\vfill\vfill

\begin{center}

{\LARGE  {\sffamily Spontaneous symmetry breaking in the \\[2mm]
non-linear Schr\"odinger hierarchy with defect}}

\vfill

{\large V. Caudrelier\footnote{caudreli@lapp.in2p3.fr}
 and E.
Ragoucy\footnote{ragoucy@lapp.in2p3.fr}\\[.21cm]
 Laboratoire de Physique Th{\'e}orique \LAP\footnote{UMR 5108
  du CNRS, associ{\'e}e {\`a} l'Universit{\'e} de
Savoie.}\\
9, chemin de Bellevue, BP 110, \\
F-74941  Annecy-le-Vieux Cedex, France.\\[.21cm]}
\end{center}

\vfill\vfill\vfill

\begin{abstract}
We introduce and solve the one-dimensional quantum non-linear
Schr\"o\-din\-ger (NLS) equation for an $N$-component field
defined on the real line with a
defect sitting at the origin. The quantum solution
is constructed using the quantum inverse scattering method based
on the concept of Reflection-Transmission (RT) algebras recently
introduced.
\\
The symmetry of the model is generated by the reflection and
transmission defect generators defining a defect
subalgebra.  We classify all the corresponding reflection and
transmission matrices. This provides the possible boundary
conditions obeyed by the canonical field and we compute these
boundary conditions explicitly.
\\
Finally, we exhibit a phenomenon of spontaneous symmetry breaking
induced by the defect and identify the unbroken generators as
well as the exact remaining symmetry.
\end{abstract}

\vfill
\centerline{PACS numbers: 02.30.Ik, 03.70.+k, 11.10.Kk}
\vfill
\rightline{\texttt{math-ph/0411022}}
\rightline{\LAP-1076/04}
\rightline{November 2004}

\newpage
\pagestyle{plain}
\setcounter{footnote}{0}

\section{Introduction}

In the framework of quantum integrable systems (qIS), a
challenging question which is addressed nowadays, is the treatment
of a localized defect (or impurity), sitting at the origin,
reflecting and transmitting particles. The ever-increasing
progress in handling quantum systems has made it possible to
create experiments able to probe the 1D behaviour of quantum gases
(see e.g. \cite{BP}).
In this context,  the construction of realistic qIS, such as ones with
defects, becomes crucial.
This is a very debated issue and essentially, two different
approaches were developed, that we briefly present.

\bigskip

Originally, qIS have been studied on the real line. Powerful
tools, such as quantum inverse scattering method, have been
developed for such a purpose. The framework is now well-known
and relies on the possibility to adapt the inverse scattering method
\cite{Mar55,GeLe,GGKM} and the arguments of Shabat and
Zakharov \cite{Zhak} to the quantum case, e.g.
\cite{FaSkTa,FaTa,Fad}. This subtle
construction can then be related to algebraic structures such as
Lie algebras, quantum groups or Yangians.
Of particular interest is the
Zamolodchikov-Faddeev (ZF) algebra \cite{ZF}.
It describes the scattering of quasi-particles and allows to
construct a hierarchy (i.e. an infinite set of Hamiltonians in
involution).  The definition of the
ZF algebra relies on an $S$-matrix, which obeys the Yang-Baxter
(YB) equation and a unitarity relation. It is interpreted as the basic two-body scattering
matrix of the quasi-particles.
This $S$-matrix is just
the usual $R$-matrix of quantum groups. It has been recently shown
that the ZF
algebra allows to construct the symmetry of this hierarchy, via
the notion of well-bred operators \cite{wellbred}. These operators realize a
quantum group defined in the FRT formalism.

As a fundamental example, the non-linear Schr\"odinger (NLS)
equation (for a review see e.g. \cite{Gut}) has played a central role
in the development of QISM, for Lax pairs and ZF algebras, e.g.
\cite{solQ1,solQ2,TWC,Gro,Dav,KS,PWZ,DavGut}.
In particular, the canonical field of the quantum NLS model has
been constructed using a ZF algebra based on the $gl(N)$-Yangian
$R$-matrix . Then, it was proved that the Yangian of
$gl(N)$ is the symmetry algebra of the NLS hierarchy \cite{yangien}, a
realization of which in terms of ZF generators being given by
well-bred operators \cite{wellbred}.

\bigskip

Later on, the study of qIS defined on the half-line was
undertaken. The pioneering work of Cherednik \cite{Cher} and
Sklyanin \cite{Skly} underlined the importance of the so-called
reflection matrix, obeying a reflection or boundary Yang-Baxter
equation (RE). This matrix encodes the reflection of particles on
the boundary, and the RE ensures the consistency between
bulk and boundary scattering.

At the algebraic level, two frameworks were developed. In the
first approach \cite{FrKo,GhoZam}, the ZF algebra is kept and a
new operator $Z_{ij}$, corresponding to the boundary, is
consistently added. From this point of view, the boundary is
assimilated to an infinitively heavy particle (an impurity),
sitting at the origin, on which the particles reflects, through
the reflection matrix. Although efficient for dealing with
correlation functions, this first approach is however unable to
treat the off-shell properties of the canonical fields. That was
the motivation of the second approach, where a (non-central)
extension of the ZF algebra, the boundary algebra \cite{boundalg}
was defined. It contains a new generator $b_{ij}(k)$ accounting
for the reflection of a particle (of impulsion $k$) on the
boundary. From this point of view, the boundary does not appear as
a heavy particle anymore, but more as a border of the half-line on
which the particles reflects. The value of the boundary generator
in a given (boundary algebra) Fock space gives back the reflection
matrix \cite{Fockbound}, making the contact with the previous
approach. In this context, the different possible reflection
matrices classify the scalar representations of the reflection
subalgebra and also the different integrable boundary conditions
obeyed by the canonical fields of the qIS defined on the
half-line. The second approach also enables  the construction of
the canonical fields of the theory.  Just like for the qIS on the
whole line, the boundary algebra allows to construct an infinite
hierarchy associated to the models on the half-line. Moreover, the
$b_{ij}(k)$ generators, which form a reflection subalgebra, appear
to be the symmetry of the hierarchy \cite{MRSbreaking}. In this
context, the reflection algebra is a subalgebra of the quantum
group which is a symmetry of the corresponding model defined on
the line \cite{ZFbound}. This second approach has been
successively applied on the fundamental examples of scalar and
$N$-component NLS model defined on the half-line \cite{GLM}.

\bigskip

In the 90's, the question of a defect which reflects and transmits
particles was studied. The approach  followed
Cherednik's original point of view to incorporate transmission.
This led to the first notion of reflection-transmission algebra,
together with reflection and transmission matrices
\cite{Delfino,Konik,Saleur}.
However, it was shown later on \cite{nogo} that the consistent
definition of these algebras together with the assumption of Lorentz invariance
implies that either the $S$-matrix
must be trivial, or there must be only reflection or only
transmission to maintain the integrability of the system. The case
of purely reflecting impurity just reproduces the half-line case
in the first approach, while the purely transmitting impurity
offers new possibilities (see for instance examples in \cite{corrigan}).

Recently, a new type of reflection-transmission (RT) algebras was
introduced \cite{MRSlet,MRS} along the line of the second approach for the
half-line. These algebras are extensions of boundary algebras and
contain, apart from ZF-like generators, two new generators
$r_{ij}(k)$ and $t_{ij}(k)$ corresponding to the reflection and
the transmission of a particle through the defect at the origin.
Contrarily to the first reflection-transmission algebras, the RT
algebras can have both reflection and transmission and a
non-trivial $S$-matrix (obeying YB and unitarity equations). This crucial point
can be related to the fact that RT algebras are more general than
the original ones. A detailed study of this fact and the proof that the
original reflection-transmission algebras are a special case of RT algebras
will be published elsewhere \cite{CMRS}. The RT algebras also
allow to construct an infinite hierarchy associated to the models
with defect, $r_{ij}(k)$ and $t_{ij}(k)$ generating the symmetry
algebra (called the defect subalgebra) of the hierarchy \cite{ZFrt}.
The defect subalgebra is also a subalgebra of the quantum group
symmetry algebra of the corresponding systems on the line
\cite{ZFrt}. The representations of the defect algebra classify the
integrable boundary conditions obeyed by the canonical fields of
the theory. RT algebras framework has been worked out by generalizing
the study of a free field on the line with a defect.
The validity of this approach has been given via the
resolution at the quantum level of the scalar NLS model with a
defect  \cite{CMRbig,CMRletter}.

\bigskip

Although convincing, the scalar NLS model possesses a scalar
$S$-matrix, so that one may wonder whether the concept of RT
algebras is still relevant for systems with internal degrees of
freedom, i.e. when the $S$-matrix is a true matrix. It is this
question that we want to address in this paper, through the study
of the quantum $N$-component NLS equation with defect. The article
is organized as follows. In section \ref{sect-RT} we briefly
review some basic notions on RT-algebras, focusing on the NLS
systems. Then, in section \ref{sect-NLS}, we show that the
RT-algebra allows to solve this system by constructing the
canonical fields of the theory. We also show that the hierarchy
constructed from the RT algebra is the one associated to the NLS
model, and that the defect subalgebra is a symmetry algebra of
this hierarchy. Section \ref{sect-class} is devoted to the
classification of the scalar representations of the defect algebra
and to the presentation of boundary conditions obeyed by the
canonical fields for each of the representations. Finally, in
section \ref{sect-break}, we show that the impurity induces a
spontaneous symmetry breaking of the defect algebra, and we
compute the unbroken symmetry algebra of the system.

\section{RT algebras \label{sect-RT}}

\subsection{Basic structures}
Let us recall briefly the ingredients involved in RT algebras
\cite{MRS}, using compact tensor notations  for
convenience. Our starting point is the following $\cs$-matrix
\begin{eqnarray}
\label{matS}
    \label{Smat}
    \cs_{12}(k_1,k_2) &=& \left(\begin{array}{cccc}
    s_{12}(k_1-k_2) & 0 & 0 & 0\\
    0 & s_{12}(k_1+k_2) & 0 & 0\\
    0 & 0 & s_{12}(-k_1-k_2) & 0\\
    0 & 0 & 0 & s_{12}(-k_1+k_2)
    \end{array}\right)\nonu
&&\mb{with} s_{12}(k)=\frac{k\,\II_N
\otimes\II_N-ig\,P_{12}}{k+ig}\,,
\end{eqnarray}
where $P_{12}$ is the usual flip operator between two auxiliary
spaces $End(\CC^{N})$. They correspond to internal degrees of
freedom. One recognizes in $s_{12}(k)$ the $S$-matrix of the NLS
equation on the line, and in the particular form of $\cs$ the one
used for solving the scalar NLS equation with defect
\cite{CMRbig}. We would like to stress that the total auxiliary
space is $End(\CC^2)\otimes End(\CC^{N})$, as it can be seen from
the $\cs$-matrix (\ref{matS}). The latter satisfies the unitarity
and Yang-Baxter equations as required
\begin{eqnarray*}
\cs_{12}(k_{1},k_{2})~\cs_{21}(k_{2},k_{1})&=&(\II_2\otimes
\II_2)\otimes (\II_N \otimes \II_N)\,,\nonu
\cs_{12}(k_{1},k_{2})~\cs_{13}(k_{1},k_{3})~\cs_{23}(k_{2},k_{3})&=&
\cs_{23}(k_{2},k_{3})~\cs_{13}(k_{1},k_{3})~\cs_{12}(k_{1},k_{2})\,.
\end{eqnarray*}
We will split the index $\alpha$ corresponding to these auxiliary
spaces into $\alpha = (\xi,i)$, with $\xi=\pm$ labelling the part
of the half-lines $\RR^\pm$ where processes take place, and
$i=1,...,N$ the internal (``isotopic") degrees of freedom.
\\
As usual, we refer to $k \in \RR$ as a spectral parameter. \\

The  RT algebra corresponding to $\cs(k_{1},k_{2})$
is an associative algebra with
identity element $\bf 1$ and two types of generators, $\{a_\alpha
(k),\, \adag_{\alpha } (k)\}$ and $\{r_\alpha^\beta (k), \,
t_\alpha^\beta (k)\}$, called bulk and defect (reflection and
transmission) generators, respectively, subject to the following
relations
\begin{eqnarray}
a_{1}(k_1)a_{2}(k_2) &=& \cs_{21}(k_{2},k_{1})
a_{2}(k_2)a_{1}(k_1)\label{rt-1}\\
\adag_{1}(k_1)\adag_{2}(k_2) &=& \adag_{2}(k_2)\adag_{1}(k_1)
\cs_{21}(k_{2},k_{1})
\label{rt-2}\\
a_{1}(k_1)\adag_{2}(k_2) &=&
\adag_{2}(k_2)\cs_{12}(k_{1},k_{2})a_{1}(k_1)+
\delta(k_1-k_2)
\Big({\bf 1} +t_{1}(k_{1})\Big)\delta_{12}\nonu && +\delta(k_1+k_2)
r_{1}(k_1)\delta_{12}\quad \label{rt-3}
\end{eqnarray}
\begin{eqnarray}
a_{1}(k_1)\,t_{2}(k_2) &=& \cs_{21}(k_{2},k_{1})\,
t_{2}(k_2)\,\cs_{12}(k_{1},k_{2})\, a_{1}(k_1)
\label{rt-4}\\
a_{1}(k_1)\,r_{2}(k_2) &=& \cs_{21}(k_{2},k_{1})\,
r_{2}(k_2)\,\cs_{12}(k_{1},-k_{2})\, a_{1}(k_1)
\label{rt-5}\\
t_{1}(k_1)\,\adag_{2} (k_2) &=& \adag_{2}(k_2)\, \cs_{12}(k_{1},k_{2})\,
t_{1}(k_1)\,\cs_{21}(k_{2},k_{1})
\label{rt-6}\\
r_{1}(k_1)\,\adag_{2} (k_2) &=& \adag_{2}(k_2)\, \cs_{12}(k_{1},k_{2})\,
r_{1}(k_1)\,\cs_{21}(k_{2},-k_{1}) \qquad \label{rt-7}
\end{eqnarray}
\begin{eqnarray}
 \!\! \!\! && \!\!\!\!
\cs_{12}(k_{1},k_{2})\, t_{1}(k_1)\,
\cs_{21}(k_{2},k_{1})\, t_{2}(k_2)\, =\, t_{2}(k_2)\,
\cs_{12}(k_{1},k_{2})\, t_{1}(k_1)\, \cs_{21}(k_{2},k_{1})
\qquad\label{rt-8}\\
 \!\! \!\! && \!\!\!\!
\cs_{12}(k_{1},k_{2})\, t_{1}(k_1)\,
\cs_{21}(k_{2},k_{1})\, r_{2}(k_2)\, =\, r_{2}(k_2)\,
\cs_{12}(k_{1},\text-k_{2})\, t_{1}(k_1)\, \cs_{21}(\text-k_{2},k_{1})
\qquad\quad\label{rt-9}\\
 \!\! \!\! \!\! && \!\!\!\!
\cs_{12}(k_{1},k_{2})\, r_{1}(k_1)\,
\cs_{21}(k_{2},\text-k_{1})\, r_{2}(k_2)\, =\, r_{2}(k_2)\,
\cs_{12}(k_{1},\text-k_{2})\, r_{1}(k_1)\, \cs_{21}(\text-k_{2},\text-k_{1})
\qquad\quad\label{rt-10}
\end{eqnarray}
where $r(k)$ and $t(k)$ satisfy:
\begin{eqnarray}
t(k)t(k)+r(k)r(-k) &=& {\bf 1} \label{rt-12a}\\
   \qquad t(k)r(k)+r(k)t(-k) &=& 0 \,.\label{rt-12b}
\end{eqnarray}
We refer to \cite{MRS} for the significance of the RT-algebras,
and to \cite{CMRbig,CMRletter} for their use in the context of
NLS.

The key point for physical applications is the possibility to
construct Fock representations of the above algebra. As explained
in \cite{MRS} and explicitly shown in \cite{CMRbig,CMRletter},
this allows to apply the quantum inverse scattering method to get
off-shell quantum fields and to compute scattering amplitudes. We
know that these representations involve a particular (vacuum)
state $\Omega$ annihilated by $a(k)$ and are uniquely determined
by the two matrices $\calr$ and $\ct$ defined by
\begin{equation}
\label{rep_rt}
r(k)\,\Omega=\calr(k)\,\Omega~~~~
\text{and}~~~~t(k)\,\Omega=\ct(k)\,\Omega\,.
\end{equation}
In the present context of NLS with defect, inspired by the
scalar case \cite{CMRletter} we take
\begin{equation}
\label{formRT}
\calr(k) = \left(\begin{array}{cc} R_+(k) & 0\\0 &
R_-(k)\end{array}\right)\, , \qquad \ct(k)=\left(\begin{array}{cc}
0 & T_+(k)\\T_-(k) & 0\end{array}\right)\, ,
\end{equation}
where $R_\pm(k),T_\pm(k)$ are now $N\times N$ matrices. We recall
that $\calr$ and $\ct$ must satisfy
\begin{eqnarray}
\calr(k)^\dagger = \calr(-k)\, , ~~ \ct(k)^\dagger = \ct(k)\,
, \label{dag}\\
\ct(k) \ct(k) + \calr(k) \calr(-k)  = \II_N \, , \\
\ct(k) \calr(k) +  \calr(k) \ct(-k) = 0 \,  , \label{unit}
\end{eqnarray}
where $^\dagger$ denotes transposition and conjugation.

\subsection{Corresponding hierarchies}\label{hierarchy}

It was shown in \cite{ZFrt} that one can naturally associate a
hierarchy of Hamiltonians to any RT algebra as follows
\begin{equation}
\wt H^{(n)}_{RT}=\int_{\RR}dk\, k^n\,
a^{\dagger\alpha}(k)a_\alpha(k),\ n\in\ZZ_{+}\, .
\end{equation}
We remind the reader that these natural Hamiltonians are not all in involution
in general
\begin{equation}
[\wt H^{(m)}_{RT},\wt H^{(n)}_{RT}]= [(-1)^m - (-1)^n] \int_{\RR}dk\,
k^{m+n}\, \, a^{\dag\alpha}(k)r_\alpha^\beta(k)a_\beta(-k) \,.
\label{comHRT}
\end{equation}
Instead, we propose to consider the Hamiltonians
\begin{equation}
H^{(n)}_{RT}=\int_{\RR}dk\, |k|^n\,
a^{\dagger\alpha}(k)a_\alpha(k),\ n\in\ZZ_{+}\, ,
\end{equation}
which are all in involution and coincide with the previous ones when
$n$ is even:
\begin{equation}
[ H^{(m)}_{RT}, H^{(n)}_{RT}]= 0 \mb{;}
H^{(2m)}_{RT}=\wt H^{(2m)}_{RT}\,.
\end{equation}
One can also compute
\begin{equation}
[ H^{(m)}_{RT}, \wt H^{(n)}_{RT}]= [1- (-1)^n] \int_{\RR}dk\,
|k|^{m}\,k^{n}\, \, a^{\dag\alpha}(k)r_\alpha^\beta(k)a_\beta(-k) \,.
\end{equation}
Interpreting $\wt H_{RT}^{(1)}$ and $\wt H_{RT}^{(2)}$ as the
momentum and the Hamiltonian of the system, (\ref{comHRT}) shows
that translation invariance is broken, as expected from the
presence of a defect at the origin. However, $H_{RT}^{(1)}$ is
conserved: the modulus of the impulsion is conserved. Remark the
noticable exception of purely transmiting systems ($r(k)=0$), where
$H_{RT}^{(1)}$ is conserved, in accordance with the above
interpretation.

The existence of an infinite number of Hamiltonians in involution has
to be related to the integrability of the models under consideration.
For instance, in \cite{CMRbig,CMRletter}, the authors showed that
$H_{RT}^{(2)}$ as defined above was the Hamiltonian of {\it
scalar} NLS with defect, the other Hamiltonians being integrals of
motion.

Finally, from (\ref{rt-4})-(\ref{rt-7}), one easily gets
\begin{equation}
\label{symmetry}
[H^{(n)}_{RT},t(k)]=0 \mb{and}
[H^{(n)}_{RT},r(k)]=0\,.
\end{equation}
This shows that the subalgebra $\cds$
(\ref{rt-8})-(\ref{rt-10}) generated by $r(k)$ and $t(k)$, which
we call the defect algebra, is the symmetry algebra
of the hierarchy described by $H_{RT}^{(n)}$. (\ref{rt-10}) shows
that the reflection algebra is itself a subalgebra
of $\cds$ and this situation should be compared to the
well-known fact that the reflection algebra is symmetry algebra of
various integrable systems on the half-line where only reflection
occurs.

\section{NLS hierarchy with defect\label{sect-NLS}}

\subsection{The $gl(N)$ model}

As in the scalar case \cite{CMRbig,CMRletter}, the quantum NLS
model with defect is described by an equation of motion on
$\RR\setminus \{0\}$ and boundary conditions at $x=0$ to be
satisfied by the field. Here, the field $\Phi_0(x,t)$ is a
$N$-component vector which splits according to
\begin{equation}
\Phi_0(x,t)=\theta(x)\Phi_{+,0}(x,t)+\theta(-x)\Phi_{-,0}(x,t)
\end{equation}
where
\begin{equation}
\Phi_{\xi,0}(x,t)=\sum_{j=1}^N \Phi_{\xi,j}(x,t)~e_j~~,~~\xi=\pm
\label{def:Phi_j}
\end{equation}
and $e_j$ is the $j$-th canonical basis vector of $\CC^N$. From
this point, using the convenient tensor notations already
introduced, the $gl(N)$ model can be dealt with in a natural way
similar to the treatment detailed in \cite{CMRbig,CMRletter}. The
fundamental ingredient is the Fock representation of the RT
algebra defined above labelled by the choice
\begin{eqnarray}
\label{coef+} R_+(k) = \frac{bk^2 + i(a-d)k + c}{bk^2 + i(a+d)k -
c}~\II_N \, , \qquad T_+(k) = \frac{2i\alpha k}{bk^2 + i(a+d)k -
c}~\II_N \, ,
 \\
\label{coef-} R_-(k) = \frac{bk^2 + i(a-d)k + c}{bk^2 - i(a+d)k -
c}~\II_N \, , \qquad T_-(k) = \frac{-2i{\overline \alpha} k}{bk^2
- i(a+d)k - c}~\II_N \, ,
\end{eqnarray}
where
\begin{equation}
\{a,...,d \in \RR,\, \alpha \in \CC\, :\, ad -bc = 1,\, {\overline
\alpha} \alpha = 1 \} \, .
\end{equation}
We further require
\begin{equation}
\left\{\begin{array}{cc} a+d-\frac{b}{|b|}\sqrt{(a-d)^2+4} \leq 0
\, ,
& \quad \mbox{$b\neq 0$}\, ,\\[1ex]
c (a+d)^{-1} \geq 0\, ,
& \quad \mbox{$b=0$}\, ,\\[1ex]
\end{array} \right.
\end{equation}
to avoid bound states since we concentrate on the scattering
theory.

Then the constructions detailed in \cite{MRS,CMRbig} provide us
with the elements $\{\ch, \cd, \Omega, \Phi\}$ necessary for a
second quantized theory:
\begin{itemize}

\item A Hilbert space $\ch$ with positive definite scalar product
$\langle \cdot\, ,\, \cdot \rangle$, which describes the states of
the system;

\item An operator valued distribution $\Phi(x,t)$, defined on a
dense domain $\cd\subset \ch$, the finite particle subspace, and
satisfying the equation of motion, the boundary conditions in mean
value on $\cd$ and the equal time canonical commutation relations;

\item A distinguished normalizable state $\Omega \in \cd$ -- the
vacuum.
\end{itemize}
In this context the off-shell field is obtained by the quantum
inverse scattering method and reads
\begin{eqnarray}
\Phi_{\xi,0}(x,t)=\sum_{n\ge 0}(-g)^n\Phi^{(n)}_{\xi,0}(x,t)
\end{eqnarray}
where
\begin{eqnarray}
\Phi^{(n)}_{\xi,0}(x,t)=\frac{1}{N^n}tr_{1\ldots
n}\left[\Phi^{(n)}_{\xi,01\ldots n}(x,t)\right]~~,~~n\ge 1
\end{eqnarray}
$tr_j$ being the trace over the $j$-$th$ auxiliary space, and
\begin{eqnarray}
\Phi^{(n)}_{\xi,01\ldots n}(x,t)=\int_{\RR^{2n+1}} \prod_{i=1\atop
j=0}^n \frac{dp_i}{2\pi}\frac{dq_j}{2\pi}\,
\adag_{\xi,1}(p_1)\ldots \adag_{\xi,n}(p_n)a_{\xi,n}(q_n)\ldots
a_{\xi,1}(q_1)a_{\xi,0}(q_0) \nonumber \\
\times\frac{e^{i\sum\limits_{j=0}^n(q_j x-q^2_j t)-i
\sum\limits_{i=1}^n(p_i x-p^2_i t)}}
{\prod\limits_{i=1}^n(p_i-q_{i-1} -\xi i\epsilon) (p_i-q_i -\xi
i\epsilon)}~~,~~n\ge 0~~~~~~~~
\end{eqnarray}
The conjugate field $\Phidag(x,t)$ is constructed in the same way
from
\begin{eqnarray}
\Phi^{\dagger(n)}_{\xi,01\ldots n}(x,t)=\int_{\RR^{2n+1}}
\prod_{i=0\atop j=1}^n \frac{dp_i}{2\pi}\frac{dq_j}{2\pi}\,
\adag_{\xi,0}(p_0)\adag_{\xi,1}(p_1)\ldots
\adag_{\xi,n}(p_n)a_{\xi,n}(q_n)\ldots
a_{\xi,1}(q_1) \nonumber \\
\times\frac{e^{i\sum\limits_{j=1}^n(q_j x-q^2_j t)-i
\sum\limits_{i=0}^n(p_i x-p^2_i t)}}
{\prod\limits_{i=1}^n(q_i-p_{i-1} +\xi i\epsilon) (q_i-p_i +\xi
i\epsilon)}~~,~~n\ge 0~~~~~~~~
\end{eqnarray}
and one can show that they satisfy the equal time canonical
commutation relations
\begin{eqnarray}
\left[\Phi_j(x,t),\Phi_k(y,t)\right] =
\left[\Phidag_j(x,t),\Phidag_k(y,t)\right]= 0 \, \\
\left[\Phi_j(x,t),\Phidag_k(y,t)\right]= \delta_{jk}~\delta(x-y)
\,,
\end{eqnarray}
where
\begin{equation}
\Phi_j(x,t)=\theta(x)\Phi_{+,j}(x,t)+\theta(-x)\Phi_{-,j}(x,t)\,,\qquad
j=1,\ldots,N\,,\qquad
\end{equation}
and $\Phi_{\pm,j}(x,t)$ has been defined in (\ref{def:Phi_j}).
 As usual for the equation of motion for quantum NLS, the
delicate point lies in the cubic term. One needs to specify a
normal ordering $:~:$ and in our case, the normal ordered cubic
term must be compatible with the use of auxiliary spaces in the
tensor notation. In this respect, we can show the following
\begin{theo}
The quantum field $\Phi_0(x,t)$ is solution of the $gl(N)$ NLS
model with defect. It obeys
\begin{equation}
\forall \varphi,\psi\in\cd,\quad
(i\partial_t+\partial_x^2)\langle\varphi,\Phi_0(x,t)\,\psi\rangle
=2g \,\langle\varphi\,,\,tr_1\, :\Phi_0\widetilde{\Phi}^\dagger_1
\Phi_1:(x,t)\,\psi\rangle
\end{equation}
and
\begin{eqnarray}
\label{limits}
\lim_{x \downarrow 0}
\left(\begin{array}{cc} \langle \varphi \, ,\, \Phi_0 (t,x) \psi \rangle  \\
\prt_x \langle \varphi \, ,\, \Phi_0 (t,x) \psi \rangle
\end{array}\right) = \alpha
\left(\begin{array}{cc} a\,\II_N & b\,\II_N\\ c\,\II_N&d\,\II_N \end{array}\right)
\lim_{x \uparrow 0}
\left(\begin{array}{cc} \langle \varphi \, ,\, \Phi_0 (t,x) \psi \rangle  \\
\prt_x \langle \varphi \, ,\, \Phi_0 (t,x) \psi \rangle
\end{array}\right) \, ,
\end{eqnarray}
\begin{eqnarray}
 \lim_{x\to \pm \infty}
\langle \varphi \, ,\, \Phi_0 (t,x) \psi \rangle = 0\,.
\end{eqnarray}
\end{theo}
\prf
The proof of this result is just a generalization to the vector
case of the proof detailed in \cite{CMRbig,CMRletter} once the
meaning of $:~:$ and $\widetilde{\Phi}^\dagger$ is given.

First,
the normal ordering is chosen in the same way as in \cite{GLM}: as
usual, all creation operators stand to the left of all
annihilation operators but with the further requirement that the
original order of the creators is preserved while that of two
annihilators is conserved if both belong to the same $\Phi$ or
$\widetilde{\Phi}^\dagger$ and inverted otherwise.

Second, for all $n\ge 1$, we define
\begin{eqnarray*}
\wt\Phi^{\dagger(n)}_{\xi,01\ldots n}(x,t)=\int_{\RR^{2n+1}}
\prod_{i=0\atop j=1}^n \frac{dp_i}{2\pi}\frac{dq_j}{2\pi}\,
\adag_{\xi,0}(p_0)\adag_{\xi,1}(p_1)\ldots
\adag_{\xi,n}(p_n)\cp_{01\ldots n}\,a_{\xi,n}(q_n)\ldots
a_{\xi,1}(q_1) \nonumber \\
\times\frac{e^{-i\sum\limits_{j=1}^n(q_j x-q^2_j t)+i
\sum\limits_{i=0}^n(p_i x-p^2_i t)}}
{\prod\limits_{i=1}^n(q_i-p_{i-1} +\xi i\epsilon) (q_i-p_i +\xi
i\epsilon)}\,~,~~~~~~~~
\end{eqnarray*}
with $\cp_{01\ldots n}=P_{01}P_{02}\cdots P_{0n}$.
Then, $\widetilde{\Phi}^\dagger$ is given by
\begin{eqnarray}
\widetilde{\Phi}^{\dagger}_{\xi,0}(x,t)= \Phi^{\dagger
(0)}_{\xi,0}(x,t)+\sum_{n\ge1}(-g)^n\frac{1}{N^n} tr_{1\ldots
n}\left[\wt\Phi^{\dagger(n)}_{\xi,01\ldots n}(x,t)\right]
\end{eqnarray}
This ensures the compatibility between the normal ordering $:~:$
and the use of auxiliary spaces to any order in the cubic term, so
that the
order by order proof \textit{\`{a} la Rosales} \cite{Gro,Dav}
works at the quantum level.
\finprf

The case treated here corresponds to "scalar  boundary conditions",
i.e. the boundary conditions do not affect the internal degree of
freedom ("isospin"). More general boundary conditions will be studied
below (see section \ref{newBound}).

For completeness, we mention that the RT algebra under
consideration is also suitable to compute transition amplitudes
between "in" and "out" states. A transition between
$|k_1,\xi_1,i_1;\ldots;k_n,\xi_n,i_n>^{in},~k_\ell\in\RR^{-\xi_\ell}$
ordered according to $|k_n|>\ldots>|k_1|$ and
$^{out}<p_1,\nu_1,j_1;\ldots;p_m,\nu_m,j_m|,~p_\ell\in\RR^{\nu_\ell}$
ordered according to $|p_1|>\ldots>|p_m|$ is computed thanks to
the identification (see \cite{CMRbig} for more details)
\begin{eqnarray*}
|k_1,\beta_1;\ldots;k_n,\beta_n>^{in}&=&\adag_{\beta_1}(k_1)\ldots
\adag_{\beta_n}(k_n)\Omega~,~~\beta_\ell=(\xi_\ell,i_\ell)\\
|p_1,\alpha_1;\ldots;p_m,\alpha_m>^{out}&=&\adag_{\alpha_1}(p_1)\ldots
\adag_{\alpha_m}(p_m)\Omega~,~~\alpha_\ell=(\nu_\ell,j_\ell)
\end{eqnarray*}
It is completely determined by the exchange relations encoded in
the RT algebra and is made out of the fundamental quantities
$\cs$, $R_\pm$ and $T_\pm$: this is the well-known factorization
of the scattering processes.

\subsection{NLS symmetry and Hamiltonian}

{From} the previous section, we see that the quantum NLS equation
with a transmitting and reflecting defect is completely solved
(in the absence of bound states). This has been done thanks to the
concept of RT algebras which substitutes the ZF algebra in the
quantum inverse inverse method when a defect is present. One
can go further with RT algebras and show that the system is
completely integrable by exhibiting a whole hierarchy of
Hamiltonians in involution. Finally, one can also identify the
symmetry algebra of the hierarchy.

Actually, the algebraic part of this has already been done in
section \ref{hierarchy} and all we have to do is to show that the
Hamiltonian of quantum NLS with defect belongs to the hierarchy
of $H^{(n)}_{RT}$ in the Fock representation labelled by
$\cs,\calr,\ct$ defined by
(\ref{matS}), (\ref{formRT}), (\ref{coef+}) and (\ref{coef-}).

It is remarkable that the results known for NLS without defect
survives in the presence of a defect: the time evolution of the
field $\Phi_{\xi,j}(x,t)$ is generated by
$H^{(2)}_{RT}=\int_{\RR}dk\, k^2\,
a^{\dagger\alpha}(k)a_\alpha(k)$ according to
\begin{equation}
\Phi_{\xi,j}(x,t)=e^{iH^{(2)}_{RT}t}~
\Phi_{\xi,j}(x,0)~e^{-iH^{(2)}_{RT}t}\,.
\end{equation}

It follows immediately from (\ref{symmetry}) that the defect
operators $r_\alpha^\beta(k),t_\alpha^\beta(k)$ generate integrals
of motion for $gl(N)$-NLS model with defect: the defect
algebra $\cds$ is the symmetry algebra of our model.

We now turn to the study of $\cds$ for the particular
representation (\ref{formRT}).

\section{Classifying NLS reflection and transmission
matrices\label{sect-class}}

\subsection{Classification}

Taking the vacuum expectation value of (\ref{rt-8})-(\ref{rt-10}),
one gets the reflection-trans\-mission equations to be satisfied by
$\calr$ and $\ct$. It is possible to classify the solutions of
these equations starting from the particular form of
(\ref{formRT}). Actually the problem reduces to solving the
reflection equation for each submatrix
$R_+(k),R_-(k),T_+(k),T_-(k)$ separately. The problem is further
constrained by additional reflection equations involving all the
possible pairs $\{R_\pm,R_\mp\}$, $\{T_\pm,T_\mp\}$,
$\{R_\pm,T_\mp\}$ and $\{R_\pm,T_\pm\}$. Finally, we will have to
take (\ref{dag})-(\ref{unit}) into account.

To complete the classification, we will use the following lemma
(proved by direct calculation, and valid whatever the $\cs$-matrix is):
\begin{lem}\label{RT-inv}
    The RT-algebra, defined by the relations (\ref{rt-1})-(\ref{rt-12b}),
    admits as automorphisms the following dilatations:
\begin{eqnarray}
r_{\eps}(k) &\to& \mu_{\eps}(k)\,r_{\eps}(k) \mb{and} t_\pm(k) \ \to\
\left(\nu_{0}(k)\right)^{\pm1}\,t_{\pm}(k)\mb{with}\eps=\pm
\end{eqnarray}
where $\nu_{0}(k)$ and $\mu_{\eps}(k)$ are functions obeying the
relations:
\begin{eqnarray}
    \mu_{\eps}(k)\,\mu_{\eps}(-k)=1\mb{and}
    \nu_{0}(-k)=\mu_{+}(-k)\,\mu_{-}(k)\,\nu_{0}(k)\,.
\end{eqnarray}
\end{lem}

{From} the classification
made in \cite{MRSbreaking} for the reflection equation, we know
that we can consider only diagonalizable and triangularizable
solutions. In our case, (\ref{dag}) rules out
the triangular case and imposes that
\begin{equation}
\label{solutionRT}
R_\pm(k)=\rho_\pm(k)\frac{\II_N+ia_\pm k\,
\GG_\pm}{1+ia_\pm k}\,,\qquad T_\pm(k)=\tau_\pm(k)\frac{\II_N+ib_\pm k\,
\FF_\pm}{1+ib_\pm k}\,,
\end{equation}
where $a_\pm,b_\pm\in\RR\cup\{\infty\}$, $\rho_{\pm}$ and
$\tau_{\pm}$ are complex functions, and $\GG_{\pm}$ and $\FF_{\pm}$
are $N\times N$ matrices obeying
$(\GG_{\pm})^2=\II_N=(\FF_{\pm})^2$. Note that the latter condition
ensures that these matrices are diagonalizable.

Now the mixed relation severely restrict the freedom in
(\ref{solutionRT}) and one finds actually
\begin{equation}
\label{formcoeff}
R_\pm(k)=\rho_\pm(k)\,M\,\frac{\II_N\pm iak\,
\EE}{1\pm ia k}\,M^{-1}\,,\qquad
T_\pm(k)=\tau_\pm(k)\,M\,\frac{\II_N\pm iak\, \EE}{1\pm
iak}\,M^{-1}\,,
\end{equation}
where $a\in\RR\cup\{\infty\}$, $\EE$ is a diagonal matrix which
squares to $\II_N$ and $M$ is a (constant) unitary diagonalization
matrix. Finally, the functions $\rho_\pm,\tau_\pm$ are constrained
by (\ref{dag})-(\ref{unit})
\begin{eqnarray}
&&\overline{\rho}_\pm(k)=\rho_\pm(-k)\,,\qquad\overline{\tau}_-(k)=\tau_+(k)\equiv
\tau_0(k)\,,\\
&&|\rho_\pm(k)|^2+|\tau_0(k)|^2=1\,,\\
&&\rho_\mp(k)\tau_\pm(k)+\rho_\pm(k)\tau_\pm(-k)=0\,.
\end{eqnarray}
The general solution to these equation takes the form:
\begin{eqnarray}
\label{param1}
\rho_\pm(k)&=&\varepsilon_\pm
~\frac{A(k)-i}{A(k)+i}~\frac{C(k)\mp i}{C(k)\pm i}~\cos\theta(k)\\
\label{param2} \tau_\pm(k)&=&
\frac{B(k)\mp i}{B(k)\pm i}~\frac{C(k)\mp i}{C(k)\pm i}~\sin\theta(k)
\end{eqnarray}
where $\varepsilon_+,\varepsilon_- =\pm1$ and $A,B,D,\theta$ are
real-valued functions obeying
$$
A(-k)=-A(k)\,,\  B(-k)=B(k)\,,\  C(-k)=-C(k)\mb{and}
\theta(-k)=-\eps_{+}\,\eps_{-}\,\theta(k)\,.
$$
Of course, when $\theta(k)=0$, one recovers the classification for
reflection matrices given in \cite{MRSbreaking}, once one has noticed
the product law
$$
\frac{A(k)-i}{A(k)+i}~\frac{C(k)-i}{C(k)+ i}=\frac{\wt A(k)-i}{\wt A(k)+i}
\mb{with} \wt A(k)= \frac{A(k)+C(k)}{1-A(k)C(k)}\,.
$$
Using the invariance given in lemma \ref{RT-inv}, one gets
 the following reduced parametrization
\begin{eqnarray}
\rho_+(k)&=&\cos\theta(k)\equiv
\frac{1-\omega(k)^2}{1+\omega(k)^2}\,,
\label{solreduite1}\\
\rho_-(k)&=&\varepsilon_+\varepsilon_-\,\cos\theta(k)\equiv
\varepsilon_+\varepsilon_-\,\frac{1-\omega(k)^2}{1+\omega(k)^2}\,,
\label{solreduite1bis}\\
\tau_\pm(k)&=& \sin\theta(k)\equiv
\frac{2\omega(k)}{1+\omega(k)^2} \mb{with}
\omega(k)=\tan\frac{\theta(k)}{2}=-\varepsilon_-\varepsilon_+\,\omega(-k)
\,.\qquad
\label{solreduite2}
\end{eqnarray}

\subsection{New boundary conditions\label{newBound}}
\subsubsection{Generalities}

When the reflection and transmission matrices take the particular
values (\ref{coef+}) and (\ref{coef-}), we know that the field
$\Phi(x,t)$ satisfy the boundary conditions (\ref{limits}). Thus,
one can think of using the previous general parametrization to
implement more general boundary conditions on the field.

Let us begin by defining the differential operator $D_A$
associated to a function $A(k)$:
\begin{equation}
(D_Af)(y)=\int dx\ \hat{A}(x)\, f(x+y)\;,\ \forall\,
f\mb{\CC-function} \label{defDA}
\end{equation}
where $\hat{A}$ is the inverse Fourier transform of $A$.
In
particular, for $A(k)=k$ and for $A(k)=a$ (constant), one has
\begin{equation}
(D_{(k)} f)(y)=-i\frac{d}{dy}f(y)\mb{and} (D_{a} f)(y)=a\,f(y)\,.
\end{equation}
This is easily extended to a matrix $A_{ij}(k),~i,j=1,\ldots,N$
\begin{equation}
(D_Af)_i(y)=\int dx\ \sum_{j=1}^N\hat{A}_{ij}(x)\, f_j(x+y)\;,\
\forall\, f_i\mb{\CC-function},~i=1,\ldots,N\,.
\label{defDmatA}
\end{equation}
Remark that, because of the property $\wh{AB}=\wh{A}*\wh{B}$, where
$*$ denotes the convolution, we have, for any matrices $A$ and $B$,
\begin{equation}
D_{AB}=D_{A}D_{B}\,.
\label{DADB}
\end{equation}
Hence, the mapping
$A(k)\to D_{A}$ is a morphism.

Now, for given reflection and transmission matrices
$R_\pm(k),T_\pm(k)$, we look for matrix-valued differential operators
such that
\begin{equation}
\label{general-limits} \lim_{x \downarrow 0}\,
\left(\begin{array}{cc} D_{u_1}\, \II_N&  0 \\ 0 &  D_{u_2}\,
\II_N
\end{array}\right)
\left(\begin{array}{c}
\langle \varphi\, ,\, \Phi(t,x) \psi \rangle\\
\prt_{x}\langle \varphi\, ,\, \Phi(t,x) \psi \rangle \end{array}
\right)=\lim_{x \uparrow 0}\,
\left(\begin{array}{cc}
D_{V_{11}} &  D_{V_{12}} \\ D_{V_{21}} &  D_{V_{22}}
\end{array}\right)
\left(\begin{array}{c}
\langle\varphi \, ,\, \Phi(t,x) \psi \rangle\\
\prt_{x}\langle \varphi\, ,\, \Phi(t,x) \psi \rangle \end{array}
\right)
\end{equation}
where the matrix-valued differential operators  $D_{V_{kl}}$, $k,l=1,2$ are
defined as in (\ref{defDmatA}), while the scalar differential
operators $D_{u_{1}}$ and $D_{u_2}$ are defined by (\ref{defDA}).
As a notation, we will use
\begin{equation}
\label{form_mat_BC}
U(k)=\begin{pmatrix} u_{1}(k)\,\II_{N} & 0 \\
0 & u_{2}(k)\,\II_{N} \end{pmatrix} \mb{and} V(k)=\begin{pmatrix}
V_{11}(k) & V_{12}(k)  \\ V_{21}(k)  & V_{22}(k)  \end{pmatrix}
\,.
\end{equation}
For convenience, we will loosely write the boundary conditions
(\ref{general-limits}) as
\begin{eqnarray}
D_{u_{1}}\Phi(+0) = D_{V_{11}}\Phi(-0) +D_{V_{12}}\,
\prt_{x}\Phi(-0)\,,
\\
D_{u_{2}}\prt_{x}\Phi(+0) = D_{V_{21}}\Phi(-0) +D_{V_{22}}\,
\prt_{x}\Phi(-0) \,,
\end{eqnarray}
keeping in mind that they must hold
for any time, although we have omitted $t$.\\

Note that matrices (\ref{form_mat_BC}) are not the most general
ones: we merely aim at finding (simple) boundary conditions for
any solution to the RT equations. We do not try to classify all
the possible boundary conditions leading to a given solution.

Following the argument of \cite{GLM}, one can write
\begin{equation}
\langle \varphi \, ,\, \Phi_i (t,x) \psi \rangle=\int
\frac{dk}{2\pi}e^{-ik^2t}\left(\Psi_{k,i}^{+,j}(x)\chi_j^+(k)
+\Psi_{k,i}^{-,j}(x)\chi_j^-(k)\right)
\end{equation}
where
\begin{equation}
\Psi_k^\pm(x)=\theta(\mp k)\left[\theta(\mp
x)T_\mp(k)e^{ikx}+\theta(\pm x)(\II_N
e^{ikx}+R_\pm(-k)e^{-ikx})\right]
\end{equation}
are the solutions of the free problem ($g=0$) and $\chi_j^\pm(k)$
are $2N$ wave-packets.

Then, the problem (\ref{general-limits}) reduces to two copies of the following
functional equations
\begin{eqnarray}
\begin{cases}
X(k)+X(-k)R_+(-k)-Y(k)T_-(k)=0\, ,\\
Y(k)+Y(-k)R_-(-k)-X(k)T_+(k)=0\, ,
\end{cases}
\end{eqnarray}
where
\begin{eqnarray}
\begin{cases}
X(k)=u_{1}(k)\II_{N}\\
Y(k)=V_{11}(k)+ikV_{12}(k)
\end{cases}\mb{or}
\begin{cases}
X(k)=ik\,u_{2}(k)\II_{N}\\
Y(k)=V_{21}(k)+ikV_{22}(k)
\end{cases}
\end{eqnarray}
In the following, we can take $\varepsilon_+=1$ without loss of
generality.

\subsubsection{Case $R_\pm(k)=\rho_\pm(k)\II_N$ and $T_\pm(k)=\tau_\pm(k)\II_N$
\label{sect:RT=1}}

In that case we can assume $u_{1}(k)=iku_{2}(k)\equiv u(k)$ and
$V(k)$ diagonal matrix with $V_{11}=v_{1}(k)\II_{N}$ and $V_{22}=v_{2}(k)\II_{N}$.
Solving the systems in
$u$, $v_{1}$ and $v_{2}$ thanks to the parametrization (\ref{param1}), (\ref{param2})
and
plugging back the result into $X,Y$, one finds a solution which can be recasted
as:
\begin{eqnarray}
U(k)&=&2(A(k)+i)(B(k)+i)(C(k)+i)\omega(k)\,\II_{2N}\,,\\
V(k)&=&(A(k)+i)(B(k)-i)(C(k)-i)\,V_{0}(k)\,,\\
V_{0}(k)&=&(\omega(k)^2+1)\begin{pmatrix}\II_{N} & 0\\
0  &\II_{N}\end{pmatrix}+\eps_{-}(\omega(k)^2-1)\begin{pmatrix}\II_{N} & 0\\
0  &-\II_{N}\end{pmatrix}\,.
\end{eqnarray}
This corresponds,  using the
property (\ref{DADB}), to the following boundary conditions:
\begin{eqnarray}
&& 2D_{f}D_{\omega}\Phi(+0) =
\left(1-\eps_{-}+(1+\eps_{-})D_{\omega}^2\right)D_{g}\Phi(-0)
\\
&& 2D_{f}D_{\omega}\prt_{x}\Phi(+0) =
\left(1+\eps_{-}+(1-\eps_{-})D_{\omega}^2\right)D_{g}\prt_{x}\Phi(-0)
\end{eqnarray}
where
\begin{equation}\begin{array}{l}
f(k)=(A(k)+i)(B(k)+i)(C(k)+i)\\
g(k)=(A(k)+i)(B(k)-i)(C(k)-i)\,.\end{array}
\label{defifg}
\end{equation}

As examples, let us use the reduced parametrization
(\ref{solreduite1})-(\ref{solreduite2}) (\ie $f=g\sim 1$) and two
particular choices for $\omega$:
 $w(k)=1$ (so that $\eps_-=-1$) or
$w(k)=k$ (then $\eps_-=1$).
They lead respectively to the following boundary conditions:
\begin{gather}
\Phi(+0) =\Phi(-0) \mb{and}
\prt_{x}\Phi(+0) = \prt_{x}\Phi(-0) \tag*{($\omega(k)=1$)}\\
 i\prt_{x}\Phi(+0) =\prt_{x}^2\Phi(-0) \mb{and}
\prt_{x}^2\Phi(+0) = i\prt_{x}\Phi(-0)\,.\tag*{($\omega(k)=k$)}
\end{gather}
Remark that the first case corresponds to a purely transmitting system
(because $\omega(k)=1$): the reduced parametrization
(\ref{solreduite1})-(\ref{solreduite2}) then leads naturally to a trivial boundary
condition (\ie no "real" defect at the origin). More involved boundary conditions can be obtained using
different matrices $U(k)$ and $V(k)$ (see for instance section
\ref{sect:RT=gle}).

\subsubsection{Case $R_\pm(k)=\rho_\pm(k)\EE$ and $T_\pm(k)=\tau_\pm(k)\EE$
\label{sect:RT=E}}

We take for $Y(k)$ the following form:
$Y(k)=y_{0}(k)\II_{N}+y_{1}(k)\EE$. We find
\begin{eqnarray}
U(k)&=&2f(k)\omega(k)\,\II_{2N}\mb{and}
V(k)=g(k)\,V_{0}(k) \mb{with}\\
V_{0}(k) &=& (\omega(k)^2+1)\begin{pmatrix}\EE & 0\\
0  &\EE\end{pmatrix}+\eps_{-}(\omega(k)^2-1)\begin{pmatrix}\II_{N} &
0\\
0  &-\II_{N}\end{pmatrix}
\end{eqnarray}
where $f(k)$ and $g(k)$ are defined in (\ref{defifg}). It
leads to the following boundary conditions:
\begin{eqnarray}
&& D_{f}D_{\omega}\Phi(+0) =
\eps_{-}\left(\frac{\II_{N}+\eps_{-}\EE}{2}D_{\omega}^2-\frac{\II_{N}-\eps_{-}\EE}{2}
\right)D_{g}\Phi(-0) \\
&& D_{f}D_{\omega}\prt_{x}\Phi(+0) =
\eps_{-}\left(\frac{\II_{N}+\eps_{-}\EE}{2}-\frac{\II_{N}-\eps_{-}\EE}{2}D_{\omega}^2
\right)D_{g}\prt_{x}\Phi(-0)
\end{eqnarray}
which shows that the defect affects the isotopic degrees of
freedom. Indeed, these boundary conditions suggest the
introduction of the "right-handed" and "left-handed" parts of
$\Phi$ defined by
\begin{equation}
\Phi_L(t,x)=\frac{\II_N-\EE}{2}~\Phi(t,x),~~\Phi_R(t,x)=\frac{\II_N+\EE}{2}~
\Phi(t,x)\,.
\end{equation}
For instance, taking the particular examples given in the previous
section, the conditions read:
\begin{gather}
\begin{cases}
\Phi_R(+0) =\Phi_{R}(t,-0)~\mb{and}~\prt_{x}\Phi_R(+0) =\prt_{x}\Phi_{R}(t,-0)\, ,\\
\Phi_L(+0) =-\Phi_{L}(t,-0)~\mb{and}~\prt_{x}\Phi_L(+0)
=-\prt_{x}\Phi_{L}(t,-0)\, .
\end{cases}\tag*{($\omega(k)=1$)}\\
\begin{cases} i\prt_{x}\Phi_R(+0)
=\prt_{x}^2\Phi_R(-0)~\mb{and}~ -i\prt_{x}^2\Phi_R(+0) =\prt_x
\Phi_R(-0)\, ,\\
i\prt_{x}\Phi_L(+0) =\Phi_L(-0)~\mb{and}~ -i\prt_{x}^2\Phi_L(+0)
=\prt_{x}^3\Phi_L(-0)\, ,
\end{cases}\tag*{($\omega(k)=k$)}
\end{gather}
One sees that the right and left-handed parts of the field do not
satisfy the same boundary conditions: the defect scatters
differently the left and right-handed parts of the field.

\subsubsection{Case $R_\pm(k)=\rho_\pm(k)\dfrac{\II_N\pm iak\,\EE}{1\pm ia k}$
and $T_\pm(k)=\tau_\pm(k)\dfrac{\II_N\pm iak\, \EE}{1\pm iak}$ \label{sect:RT=gle}}

Again, one can take $Y(k)=y_{0}(k)\II_{N}+y_{1}(k)\EE$ to find a
solution of the form
\begin{eqnarray}
U(k)&=&2(1+iak)f(k)\omega(k)\,\II_{2N}\mb{and} V(k)=g(k)\,V_{0}(k)
\mb{with}\nonu
V_{0}(k) &=& (\omega(k)^2+1)\begin{pmatrix}\II_{N} +aki\,\EE &0\\
ik\II_{N}
&aik\,\EE\end{pmatrix}+\eps_{-}(\omega(k)^2-1)\begin{pmatrix}(1-iak)\II_{N}
&
0\\
-ik\II_{N}  &aik\,\II_{N}\end{pmatrix} \nonumber
\end{eqnarray}
This corresponds to the following boundary conditions:
\begin{eqnarray*}
D_{f}D_{\omega}(1+a\prt_{x})\Phi(+0)  &=&
\left(\frac{1-\eps_-}{2}+\frac{1+\eps_-}{2}D_{\omega}^2\right)D_{g}\Phi(-0)\nonu
&&+a~\eps_-
\left(\frac{\II_N+\eps_-\EE}{2}-\frac{\II_N-\eps_-\EE}{2}D_{\omega}^2
\right)D_{g}\prt_{x}\Phi(-0)\\
D_{f}D_{\omega}(1+a\prt_{x})\prt_{x}\Phi(+0)  &=&
\left(\frac{1+\eps_-}{2}+\frac{1-\eps_-}{2}D_{\omega}^2\right)D_{g}\prt_{x}\Phi(-0)\nonu
&&-a~\eps_-
\left(\frac{\II_N-\eps_-\EE}{2}-\frac{\II_N+\eps_-\EE}{2}D_{\omega}^2
\right)D_{g}\prt^2_{x}\Phi(-0)
\end{eqnarray*}
Taking once more the two particular examples, we get
\begin{gather}
\begin{cases}
(1+a\prt_{x})\Phi(+0)=\Phi(-0)+a\prt_{x}\left(\Phi_R(-0)-\Phi_L(-0)\right)\\
(1+a\prt_{x})\prt_{x}\Phi(+0)=\prt_{x}\Phi(-0)+a\prt^2_{x}\left(\Phi_R(-0)-\Phi_L(-0)\right)
\end{cases}\tag*{($\omega(k)=1$)}\\[2mm]
\begin{cases}
i(1+a\prt_{x})\prt_{x}\Phi(+0)=\prt^2_{x}\Phi(-0)-a\prt_{x}\Phi_R(-0)-a\prt^3_{x}\Phi_L(-0)\\
i(1+a\prt_{x})\prt^2_{x}\Phi(+0)=-\prt_{x}\Phi(-0)+a\prt^2_{x}\Phi_L(-0)+a\prt^4_{x}\Phi_R(-0)
\end{cases}\tag*{($\omega(k)=k$)}
\end{gather}
Note that the first case corresponds to a purely transmitting
system with non-trivial boundary conditions (for $a\neq 0$).
Again, the defect is sensitive to the "isospin".

\section{Spontaneous symmetry breaking\label{sect-break}}

\subsection{General discussion}
It is known \cite{MRS} that the matrices $\calr, \ct$ in
(\ref{rep_rt}) fully determine the different Fock representations
of the RT algebra. For convenience we denote by $\langle~\rangle$
the vacuum expectation value of any RT generator on $\Omega$ and
by construction one has
\begin{equation}
\langle t(k) \rangle=\ct(k),~~\langle r(k) \rangle=\calr(k)\,.
\end{equation}
Thus, under an expansion in power of $k^{-1}$, one can identify
those generators of the defect algebra whose vacuum expectation
value is nonzero. As we showed, the defect algebra is the
symmetry of the NLS hierarchy so that this mechanism is
 a spontaneous symmetry breaking of the defect algebra.
To make this more quantitative, let us introduce some notations.
We write series expansions for the generators $r_\pm(k)$,
$t_\pm(k)$ of the defect algebra as follows
\begin{equation}
r_\pm(k)=r_\pm^{(0)}+\sum_{n=1}^\infty r_\pm^{(n)}
k^{-n},~~t_\pm(k)=t_\pm^{(0)}+\sum_{n=1}^\infty t_\pm^{(n)}
k^{-n}\, .
\end{equation}
The generators corresponding to the physical symmetries are
$r_\pm^{(n)},t_\pm^{(n)},~n\ge 1$. Thus, there is spontaneous
symmetry breaking if $R^\pm(k),T^\pm(k)$ are not constant
matrices. The problem then is to identify the remaining exact
symmetry \ie one has to find the unbroken generators and the
algebra they satisfy.
In other words, we look for  generators
$\widehat{r}_\pm^{(n)},\widehat{t}_\pm^{(n)}$, for some $n\geq1$,  such that
\begin{equation}
\label{unbroken}
\langle\widehat{r}^{(n)}_\pm\rangle=0,~~\langle\widehat{t}_\pm^{(n)}\rangle=0\,.
\end{equation}
This is the question we address in the case
of NLS in the next paragraph.

\subsection{Symmetry breaking for NLS}

In this case, we know the general form of $R_\pm(k),T_\pm(k)$:
\begin{eqnarray*}
R_\pm(k)=\rho_\pm(k)\,M\,\Lambda(\pm k)\,M^{-1},
~~T_\pm(k)=\tau_\pm(k)\,M\,\Lambda(\pm k)\,M^{-1},
~~\Lambda(k)=\frac{\II_N+iak\, \EE}{1+ ia k}\,.
\end{eqnarray*}
As these are not constant matrices in general, the corresponding
symmetry is broken.
To identify the remaining symmetry algebra, we introduce
\begin{equation}
\lambda(k)=\frac{1}{2}\left[
\left(1+\frac{1-iak}{\sqrt{1+(ak)^2}}\right)\II_N+
\left(1-\frac{1-iak}{\sqrt{1+(ak)^2}}\right)\EE\right]\, ,
\end{equation}
which obeys
\begin{equation}
\lambda(k)^2=\Lambda(k)~~\text{and}~~\lambda(k)\lambda(-k)=\II_N\,.
\end{equation}
We also introduce the  generators
\begin{equation}
\label{tilde} \widetilde{r}_\pm(k)=\lambda(\mp
k)~M^{-1}\,r_\pm(k)\,M~\lambda(\mp k),~~
\widetilde{t}_\pm(k)=\lambda(\mp
k)~M^{-1}\,t_\pm(k)\,M^{-1}~\lambda(\mp k)
\end{equation}
 which we gather into
\begin{equation}
\widetilde{r}(k)= \left(\begin{array}{cc} \widetilde{r}_+(k) &
0\\0 & \widetilde{r}_-(k)\end{array}\right)\, , \qquad
\widetilde{t}(k)=\left(\begin{array}{cc} 0 &
\widetilde{t}_+(k)\\\widetilde{t}_-(k) & 0\end{array}\right)\, .
\end{equation}
After some algebra,
one finds that the new generators satisfy the following relations:
\begin{eqnarray*}
\widetilde{\cs}_{12}(k_{1},k_{2})\,
\widetilde{t}_{1}(k_1)\, \widetilde{\cs}_{21}(k_{2},k_{1})\,
\widetilde{t}_{2}(k_2) &=& \widetilde{t}_{2}(k_2)\,
\widetilde{\cs}_{12}(k_{1},k_{2})\, \widetilde{t}_{1}(k_1)\,
\widetilde{\cs}_{21}(k_{2},k_{1})
\\
 \widetilde{\cs}_{12}(k_{1},k_{2})\,
\widetilde{t}_{1}(k_1)\, \widetilde{\cs}_{21}(k_{2},k_{1})\,
\widetilde{r}_{2}(k_2) &=& \widetilde{r}_{2}(k_2)\,
\widetilde{\cs}_{12}(k_{1},-k_{2})\, \widetilde{t}_{1}(k_1)\,
\widetilde{\cs}_{21}(-k_{2},k_{1})
\\
\widetilde{\cs}_{12}(k_{1},k_{2})\,
\widetilde{r}_{1}(k_1)\, \widetilde{\cs}_{21}(k_{2},-k_{1})\,
\widetilde{r}_{2}(k_2) &=& \widetilde{r}_{2}(k_2)\,
\widetilde{\cs}_{12}(k_{1},-k_{2})\, \widetilde{r}_{1}(k_1)\,
\widetilde{\cs}_{21}(-k_{2},-k_{1}) \qquad\quad
\\
\widetilde t(k)\,\widetilde t(k)+\widetilde r(k)\,\widetilde r(-k) = {\bf 1}
&\mbox{and}&
\widetilde t(k)\,\widetilde r(k)+\widetilde r(k)\,\widetilde t(-k)
   = 0
\end{eqnarray*}
where the new matrix $\widetilde\cs$ defined by
\begin{eqnarray}
    \widetilde{\cs}_{12}(k_1,k_2) &=& \left(\begin{array}{cccc}
    \widetilde{s}_{12}(k_1,k_2) & 0 & 0 & 0\\
    0 & \widetilde{s}_{12}(k_1,-k_2) & 0 & 0\\
    0 & 0 & \widetilde{s}_{12}(-k_1,k_2) & 0\\
    0 & 0 & 0 & \widetilde{s}_{12}(-k_1,-k_2)
    \end{array}\right)\qquad\\
\widetilde{s}_{12}(k_1,k_2) &=& \lambda_1(-k_1)\,\lambda_2(-k_2)\,
s_{12}(k_1-k_2)\,\lambda_1(k_1)\,\lambda_2(k_2)\,
\end{eqnarray}
satisfies the unitarity and Yang-Baxter equations
\begin{eqnarray*}
\widetilde{\cs}_{12}(k_{1},k_{2})~\widetilde{\cs}_{21}(k_{2},k_{1})
&=&(\II_2\otimes\II_2)\otimes (\II_N \otimes \II_N)\nonu
\widetilde{\cs}_{12}(k_{1},k_{2})~\widetilde{\cs}_{13}(k_{1},k_{3})~\widetilde{\cs}_{23}(k_{2},k_{3})&=&
\widetilde{\cs}_{23}(k_{2},k_{3})~\widetilde{\cs}_{13}(k_{1},k_{3})~\widetilde{\cs}_{12}(k_{1},k_{2})\,.
\end{eqnarray*}
Thus, the generators $\widetilde r(k)$ and $\widetilde t(k)$, which by
construction
belong to the original defect algebra $\cds$, generates a
subalgebra. This subalgebra appears to be itself a
{\cdst}  defect algebra\footnote{Strictly speaking, we have shown that
 $\widetilde r(k)$ and $\widetilde t(k)$ generates a subalgebra of
 $\cdst$.}.

Computing the vacuum expectation value of these generators, we get
\begin{equation}
\langle\widetilde{r}_\pm{(k)}\rangle=\rho_{\pm}(k)\,\II_{N},~~
\langle\widetilde{t}_\pm{(k)}\rangle=\tau_{\pm}(k)\,\II_{N}\,,
\label{vactilde}
\end{equation}
so that all the off-diagonal terms of $\widetilde r_\pm(k)$ and $\widetilde t_\pm(k)$
remain unbroken. Indeed, the form (\ref{vactilde}) shows that only the
generators obtained from the $k^{-1}$ expansion of the
traces tr$(\widetilde r_\pm(k))$ and tr$(\widetilde t_\pm(k))$ are possibly
broken, depending on the exact form
of the functions $\rho_{\pm}(k)$ and $\tau_{\pm}(k)$.
In other words, the use of the generators $\widetilde r_\pm(k)$ and $\widetilde t_\pm(k)$
takes care of the symmetry breaking induced by the matrix
$M\,\Lambda(k)\,M^{-1}$,
while the expansion in $k^{-1}$ of the functions $\rho_{\pm}(k)$ and $\tau_{\pm}(k)$
will induce a "scalar-like" symmetry breaking (i.e. a breaking of
 the type $gl(N)\to sl(N)$). These two points will be illustrated in
 two examples below.

We want to stress that even if one starts with an exchange matrix
$S(k)$ depending only on the difference of the parameters, the new
matrix $\widetilde{S}$ (which defines the unbroken symmetry algebra)
depends \textit{separately} on $k_1$,
$k_2$. This shows (once again) that a reflection-transmission
algebra is naturally (although not compulsorily) associated with
an exchange matrix depending on $k_1$ and $k_2$ separately.

Let us note finally that this study completes the arguments
developed in \cite{MRSbreaking} for NLS on the half-line. Indeed,
we know that the reflection algebra is particular case of the
defect algebra when one considers $t(k)=0$. Performing the same
calculations as above, we deduce that in the case of spontaneous
symmetry breaking for NLS on the half-line, the remaining exact
symmetry is again a reflection algebra with the exchange matrix
$\widetilde{S}(k)$ and unbroken generator $\widetilde{r}(k)$.

\subsubsection{Example: $R_\pm(k)=\cos(\theta_{0}/k)\,\II_{N}$
and $T_\pm(k)=\sin(\theta_{0}/k)\,\II_N$\label{ex1}} 
In that case, we have
$\widetilde{r}(k)=r(k)$ and $\widetilde{t}(k)=t(k)$, and the
symmetry breaking is induced by the series expansion of the sine and
cosine functions. One gets
\begin{eqnarray}
    <r_{\pm}^{(2n)}>=(-1)^n\,\theta_{0}^{2n}\,\II_{N}\quad;\quad
    <r_{\pm}^{(2n+1)}>=0\\
    <t_{\pm}^{(2n+1)}>=(-1)^n\,\theta_{0}^{2n+1}\,\II_{N}\quad;\quad
    <t_{\pm}^{(2n)}>=0
\end{eqnarray}
which allow to identify directly the unbroken generators. 
Let us define
\begin{equation}
r_\pm^{(n)}=\sum_{i,j=1}^N~r_\pm^{(n),ij}~E_{ij} \mb{and}
t_\pm^{(n)}=\sum_{i,j=1}^N~t_\pm^{(n),ij}~E_{ij}\,.
\end{equation}
One easily sees that all the generators 
$r_\pm^{(n),ij}$ and $t_\pm^{(n),ij}$ with $1\leq i\neq j\leq N$ have a vanishing
vacuum expectation value, and hence remain unbroken. It is also the
case of the generators $r_\pm^{(2n+1),ii}$ and $t_\pm^{(2n),ii}$.  
Other unbroken generators are given by the
combinations $r_\pm^{(2n),ii}-r_\pm^{(2n),jj}$ and
$t_\pm^{(2n+1),ii}-t_\pm^{(2n+1),jj}$, in accordance with the
expected "scalar-like"  symmetry breaking: only the traces
$\sum_{i=1}^{N}r_\pm^{(2n),ii}$ and $\sum_{i=1}^{N}t_\pm^{(2n+1),ii}$ 
are broken.

This fact is easily extended
to any functions of $k^{-1}$ (instead of cosine and
sine), as long as they are in the class of functions
(\ref{param1})-(\ref{param2}).

\subsubsection{Example: $R_\pm(k)=\pm\cos(\theta_{0})\,M\,\Lambda(\pm k)\,M^{-1}$
; $T_\pm(k)=\sin(\theta_{0})\,M\,\Lambda(\pm k)\,M^{-1}$}
We consider the case $N=2$ and we take the following forms
\begin{eqnarray}
    \Lambda(k) &=& \left(\begin{array}{cc} 1 & 0 \\ 0 &\beta(k)
\end{array}\right) \mb{with} \beta(k)=\frac{1-iak}{1+ ia k} \\
   M &=& \left(\begin{array}{cc} \cos(\mu) & \sin(\mu) \\ -\sin(\mu) &\cos(\mu)
\end{array}\right) \mb{with} \mu\in [0,\pi[
\end{eqnarray}
A direct computation shows that
\begin{equation}
\langle r_\pm(k)\rangle= \pm \cos\theta_0 ~\Gamma(k)~~,~~\langle
t_\pm(k)\rangle=  \sin\theta_0 ~\Gamma(k)
\end{equation}
with
\begin{equation}
\Gamma(k)=\left(\begin{array}{cc} \cos^2\mu
 + \beta(k)\,\sin^2\mu & (\beta(k)-1)\,\cos\mu~\sin\mu \\
(\beta(k)-1)\,\cos\mu~\sin\mu &\sin^2\mu
 + \beta(k)\,\cos^2\mu
\end{array}\right)\,.
\end{equation}
Upon expanding the elements of $\Gamma(k)$ in powers of $k^{-1}$,
one sees that all the generators $r_\pm^{(n),ij}$,
$t_\pm^{(n),ij}$, $n>0$,
get non-vanishing vacuum expectation values (for $\mu\neq 0$ and
$\mu \neq \pi/2$):
\begin{eqnarray}
\langle r_\pm^{(n),11} \rangle&=&\pm \cos\theta_0
(-2\sin^2\mu)\left(\frac{\pm i}{a}\right)^n\\
\langle r_\pm^{(n),22} \rangle&=&\pm \cos\theta_0
(-2\cos^2\mu)\left(\frac{\pm i}{a}\right)^n\\
\langle r_\pm^{(n),12} \rangle&=&\pm \cos\theta_0
(-2\sin\mu\cos\mu)\left(\frac{\pm i}{a}\right)^n=\langle
r_\pm^{(n),21} \rangle
\end{eqnarray}
and similar expressions for $t_\pm^{(n),ij}$ replacing $\pm
\cos\theta_0$ by $\sin\theta_0$.

However, the following combinations, dictated by (\ref{tilde}),
produce  unbroken generators
\begin{eqnarray}
\widetilde{x}_\pm^{11}(k)&=&
x_\pm^{11}\,\cos^2\mu-(x_\pm^{12}+x_\pm^{21})\,\sin\mu\cos\mu
+x_\pm^{22}\,\sin^2\mu
\\
\widetilde{x}_\pm^{12}(k)&=&\frac{1\pm iak}{\sqrt{1+a^2k^2}}\left\{
x_\pm^{12}\,\cos^2\mu+(x_\pm^{11}-x_\pm^{22})\,\sin\mu\cos\mu
-x_\pm^{21}\,\sin^2\mu \right\}=\widetilde{x}_\pm^{21}(k)\nonu
\widetilde{x}_\pm^{22}(k)&=&\frac{(1\pm iak)^2}{1+a^2k^2}\left\{
x_\pm^{11}\,\sin^2\mu+(x_\pm^{12}+x_\pm^{21})\,\sin\mu\cos\mu
+x_\pm^{22}\,\cos^2\mu \right\}
\end{eqnarray}
where $x_\pm$ denote either $r_\pm(k)$ or
$t_\pm(k)$.

Indeed, a direct computation yields
\begin{eqnarray}
\langle\widetilde{r}_\pm^{11}(k)\rangle=\langle\widetilde{r}_\pm^{22}(k)\rangle=\pm
\cos\theta_0~~&,&~~
\langle\widetilde{t}_\pm^{11}(k)\rangle=\langle\widetilde{t}_\pm^{22}(k)\rangle=\sin\theta_0\\
\langle\widetilde{r}_\pm^{12}(k)\rangle=\langle\widetilde{r}_\pm^{21}(k)\rangle=0~~&,&~~
\langle\widetilde{t}_\pm^{12}(k)\rangle=\langle\widetilde{t}_\pm^{21}(k)\rangle=0
\end{eqnarray}
as expected. Similar calculations can be done when $N$ (the size of
the matrices) is greater than 2. 

Finally, let us note that in this example, we have considered constant
functions $\rho_{\pm}$ and $\tau_{\pm}$: in the more general case
where they do depend on $k$, the generators
$\widetilde{x}_\pm(k)$ will undergo a scalar-like symmetry
breaking, as in the previous example. Thus, to get true unbroken generators, a
second step (similar to that of section \ref{ex1}) is needed.

\end{document}